\documentclass[10pt]{article}
\usepackage{amsthm,amsmath,amssymb}
\usepackage{subfig,sidecap,caption}
\usepackage{paralist}
\usepackage[usenames,dvipsnames]{color}
\usepackage[pdftex,breaklinks,colorlinks,
    citecolor={BlueViolet}, linkcolor={Blue},urlcolor=Maroon]{hyperref}  
\usepackage{tikz,pgfkeys}
\usepackage{tkz-graph}
\usetikzlibrary{decorations.pathmorphing}
\usetikzlibrary{quotes}
\usetikzlibrary{arrows.meta}
\usetikzlibrary{arrows,shapes}
\usetikzlibrary{decorations.pathreplacing, decorations.shapes}
\usepackage{graphicx}
\usepackage{charter,eulervm}%
\usepackage{multirow,booktabs,array}
\usepackage{tabularx}
\usepackage[final,expansion=alltext,protrusion=true]{microtype}
\usepackage{float}
\usepackage{enumerate}
\usepackage{algorithm}         
\usepackage{algpseudocode}

\theoremstyle{plain} 
\newtheorem{theorem}{Theorem}[section]
\newtheorem{lemma}[theorem]{Lemma}

\theoremstyle{definition}

\newcommand{\hsc}[1]{{\footnotesize\sf\MakeUppercase{#1}}}
\newcommand{\problem}[2]{\hsc{#1} $\rightarrow$ \hsc{#2}}
\newcommand{\lp}[1]{\ensuremath{{\mathtt{lp}(#1)}}}
\newcommand{\rp}[1]{\ensuremath{{\mathtt{rp}(#1)}}}
\newcommand{\comment}[1]{\textbackslash\!\!\textbackslash {\em #1}}
\tikzstyle{class} = [shape=rectangle, rounded corners, draw, align=center, top color=white, bottom color=blue!20]
\tikzstyle{vertex}  = [{fill=blue,circle,draw,inner sep=1pt}]

\title{Graph Isomorphism by Conversion to Chordal (6, 3) Graphs}

\author{M. Delacorte}

\date{}

\begin{document}
\maketitle

\begin{abstract}
 Babel has shown that for an extended class of chordal (6, 3) graphs the coarsest regular simplicial partition is equivalent to the graph's automorphism partition. We give a reversible transformation for any graph to one of these graph by using Booth's reduction of a graph to a chordal graph and elimination of Babel's  forbidden subgraphs for these graphs by adding edges to them.
\end{abstract}

\section{Introduction}\label{sec:intro}
Chordal graphs are graphs where every cycle of size 4 or greater in the graph is chorded.  In a (q, t) graph [2] any set of q or less vertices will induce at most t P4s (where P4 denotes a chordless path on four vertices).  Given a graph $G = (V, E)$ let $V$ be its set of vertices, $E$ its edge set, $n = |V|$, $m = |E|$.  For $x \in V$ let $N(x) = \{y \in V:xy \in E\}$ be the set of neighbours of $x$.  If $x \cup N(x)$ induces the only clique of $G$ that contains $x$ then $x$ is a simplicial vertex.  For chordal graphs every induced subgraph has a least one simplicial vertex.  An ordered partition $V = S_1 \cup S_2 \cup... S_q$ of $V$ is simplicial $iff$ for $1 < i < q$ all the nodes in $S_i$ are simplicial nodes of the subgraph $G(S_i \cup S_{i+l} ... \cup S_q)$.  The sets $S_i$ are called cells.  Two cells $S_i$ and $S_j$ are called adjacent $iff$ at least one node of $S_i$ is adjacent to at least one node in $S_j$; they are called totally adjacent iff all node of $S_i$ are adjacent to all nodes in $Sj$.  The coarsest simplicial partion consists of $S_i$ such that $S_i$ equals the set of all simplicial nodes in $G(S_i \cup...\cup S_q)$.  A simplicial partition $S$ of $G$ is called regular $iff$ for any cells $S_i$, $S_j$ in the partition and for any $x$, $x'$ in $S_i$ we have $|N_j(x)| = |N_j(x')|$.  For extended discussions of simplicial and regular partitions see [3, 4].\par 
 Babel gives a set of five 6 vertex forbidden graphs [1] which can be used to characterizes an extended class of chordal (6, 3) graphs (see fig. 1).  He also characterizes three families of graphs stars, thin leg spiders and thick leg spiders for these graphs (see fig. 2). 

A $star$ graph.  The vertex set $V$ can be partitioned into non empty sets $K_0, K_1 , . . . , K_r, r \geq 1$, such that

\begin{enumerate}[(i)]
\item $|K_i| = |K_j| for 1 \leq i,j \leq r.$
\item There are no edges between different sets $K_i$ and $K_j$, $1 \leq i, j \leq r.$
\item $K_0 \cup K_i, 1 \leq i \leq r,$ induces a clique.
$K_0$ is the $center$ of the star and, obviously, induces a complete subgraph.
\end{enumerate} 
 
The spider graphs.  The vertex set V can be partitioned into sets $K, S$ such that

\begin{enumerate}[(i)]
\item $|K| = |S| \geq 2$, $K$ induces a complete subgraph, $S$ is a stable set.
\item There exists a bijection $f:S -> K$ such that either
\begin{enumerate}[(a)]
\item for all $s \in S, k \in K: sk \in E \leftrightarrow f(s) = k$ or
\item for all $s \in S, k \in K: sk \notin E \leftrightarrow f(s) = k.$
\end{enumerate}
\end{enumerate}
If the first of the two alternatives holds then G is said to be a $spider$ with thin
legs, otherwise with thick legs. A $P_4$ is considered to be a spider with thin legs.
Obviously, the complement of a spider with thin legs is a spider with thick legs
and vice versa.  $K$ is the center of the spider.

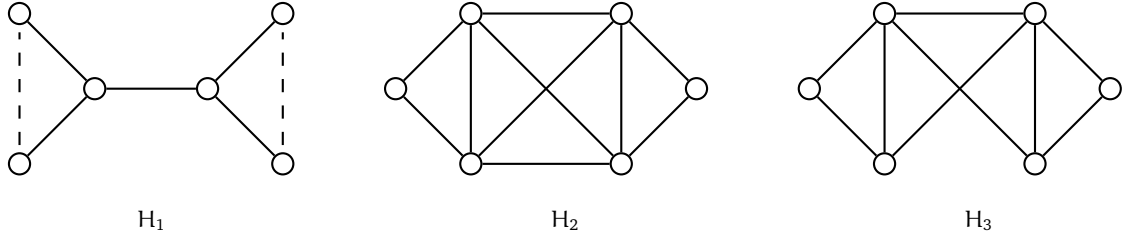
\begin{figure}[H]
	\centering\small
	
	\begin{tikzpicture}[thick, scale=0.5]
	
	\tikzstyle{every node}=[circle,draw=black,fill=white,thick,minimum size=8pt,
	inner sep=0pt]
	\draw (1,3) node (1)  {};
	\draw (1,7) node (2)  {};
	\draw (3,5) node (3)  {};
	\draw (6,5) node (4)  {};
	\draw (8,3) node (5)  {};
	\draw (8,7) node (6)  {};
	
	\draw [thick,dash pattern={on 5pt off 5pt}] (1,3.5) -- (1,6.5);
	\draw [thick,dash pattern={on 5pt off 5pt}] (8,3.5) -- (8,6.5);
	
	\draw (1)--(3)--(2);
	\draw (3)--(4)--(5);
	\draw (4)--(6);

	\draw (11,5) node (1)  {};
	\draw (13,3) node (2)  {};
	\draw (13,7) node (3)  {};
	\draw (17,3) node (4)  {};
	\draw (17,7) node (5)  {};
	\draw (19,5) node (6)  {};

	\draw (3)--(1)--(2)--(3)--(5)--(4)--(6)--(5);
	\draw (3)--(4);
	\draw (4)--(2)--(5);

	\draw (22,5) node (1b)  {};
	\draw (24,3) node (2b)  {};
	\draw (24,7) node (3b)  {};
	\draw (28,3) node (4b)  {};
	\draw (28,7) node (5b)  {};
	\draw (30,5) node (6b)  {};

	\draw (3b)--(1b)--(2b)--(3b)--(5b)--(4b)--(6b)--(5b);
	\draw (3b)--(4b);
	\draw (2b)--(5b); 
	
	\coordinate[label=left:{$H_1$}] (h1) at (5,1.5);
	
	\coordinate[label=left:{$H_2$}] (h2) at (16,1.5);
	
	\coordinate[label=left:{$H_3$}] (h3) at (27,1.5);

	\end{tikzpicture}
	
	\caption{Forbidden subgraphs for an extended class of chordal (6, 3) graphs.  Dashed edges in $H_1$ graph may or may not be present}
	\label{fig:overview}
\end{figure}

\begin{figure}[H]
	\centering\small
	
	\begin{tikzpicture}[thick, scale=0.5]
	
	
	\tikzstyle{every node}=[circle,draw=black,fill=white,thick,minimum size=8pt,
	inner sep=0pt]
	\draw (25,3) node (1)  {};
	\draw (23.5,5) node (2)  {};
	\draw (26.5,5) node (3)  {};
	\draw (22,7) node (4)  {};
	\draw (25,7) node (5)  {};
	\draw (28,7) node (6)  {};
	
	\draw (1)--(2)--(4)--(5)--(6)--(3)--(5)--(2)--(3)--(1);

	\draw (13,3) node (1)  {};
	\draw (19,3) node (2)  {};
	\draw (14.5,4.5) node (3)  {};
	\draw (17.5,4.5) node (4)  {};
	\draw (16,7) node (5)  {};
	\draw (16,9) node (6)  {};

	\draw (1)--(3)--(4)--(5)--(3);
	\draw (2)--(4);
	\draw (5)--(6);

	\draw (6,2.5) node (1)  {};
	\draw (7,2.5) node (2)  {};
	\draw (4,5) node (3)  {};
	\draw (9,5) node (4)  {};
	\draw (6.5,5.5) node (5)  {};
	\draw (4,6) node (6)  {};
	\draw (9,6) node (7)  {};
	\draw (6,8.5) node (8)  {};
	\draw (7,8.5) node (9)  {};

	\draw (5)--(1)--(2)--(5)--(3)--(6)--(5)--(4)--(7)--(5)--(9)--(8)--(5);

	\coordinate[label=left:{star}] (star) at (7.5,1.5);
	
	\coordinate[label=left:{spider with thin legs}] (thin) at (20,1.5);
	
	\coordinate[label=left:{spider with thick legs}] (thick) at (29,1.5);

	\end{tikzpicture}
	
	\caption{Examples of star and spider graphs.  Graphs used in proving the automorphism partition of an extended class of chordal (6, 3) graphs coincides with the coarsest regular simplicial partition.}
	\label{fig:overview}
\end{figure}
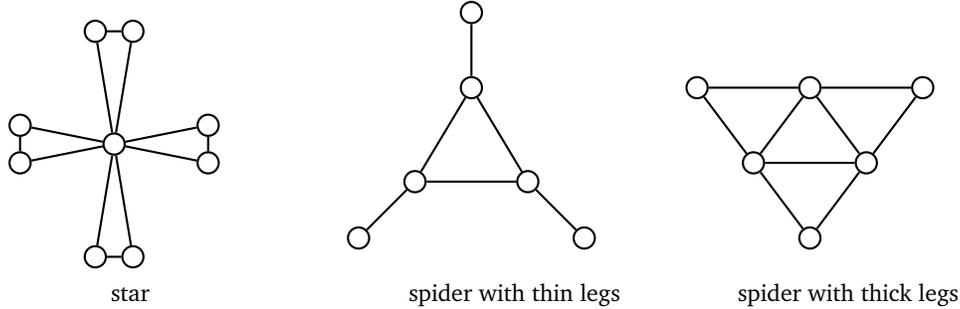

The following lemmas for a regular simplicial partition of an extended class of chordal (6, 3) graphs are proved. 

\begin{lemma}
Let $S_i, S_j$ be adjacent cells with $j > i$. Then $G(S_i \cup S_j)$ is the disjoint union of isomorphic stars or isomorphic spiders. $S_j$  consists exactly of the centers (which are of equal size).
\end{lemma}

Given a spider in $G(S_i \cup S_j)$, we now denote the vertices of the center by $w_1,w_2,..., w_t$ and the remaining vertices by $u_l,u_2,...,u_t$ in such a way that $u_lw_l \in E$ for a spider with thin legs resp. $u_lw_l \notin E$ for a spider with thick legs,
$l= 1,2,...,t$ .

\begin{lemma}
Let $G(S_i \cup S_j)$, $j > i$, be the disjoint union of spiders and let $S_k$ be adjacent to $S_i, k > i$, $k \neq j$. Then $N_k(u_l) = N_k(w_l)$  holds for all $l$.
\end{lemma}

\begin{lemma}
Let $G(S_i \cup S_k)$, $k > i$, be the disjoint union of stars for all cells $S_k$ which are adjacent to $S_i$. Further let $j$ be minimal, $j > i$, such that $S_j$ is adjacent to $S_i$  and $ k \neq j$.  Then  $N_k(u) = N_k(w)$  holds for all $u \in S_i$, $w \in S_j$  with $uw \in E.$
\end{lemma}

 Finally using the above lemmas it is proved that the coursest regular simplicial partition of an extended class of chordal (6, 3) graph is the same as its automorphism partition.
 
 \section{Preliminaries}\label{sec:preliminaries}

Before giving the algorithm to convert a graph to an extended class of chordal (6, 3) graph we give a few preliminaries.  Graph isomorphism is polynomial time reducible to chordal graph isomorphism [4] as follows map given graph G to its subdivision graph, and then connect all the vertices which were vertices in G to each other.  To transform a chordal graph to an extended class of chordal (6, 3) graph edges are added to each of the forbidden subgraphs in the graph (see fig. 3).

 \begin{figure}[H]
	\centering\small
	
	\begin{tikzpicture}[thick, scale=0.5]
	
	\tikzstyle{every node}=[circle,draw=black,fill=white,thick,minimum size=8pt,
	inner sep=0pt]
	\draw (1,3) node (1)  {};
	\draw (1,7) node (2)  {};
	\draw (3,5) node (3)  {};
	\draw (6,5) node (4)  {};
	\draw (8,3) node (5)  {};
	\draw (8,7) node (6)  {};
	
	\draw [thick,dash pattern={on 5pt off 5pt}] (1,3.5) -- (1,6.5);
	\draw [thick,dash pattern={on 5pt off 5pt}] (8,3.5) -- (8,6.5);
	
	\draw (1)--(3)--(2);
	\draw (3)--(4)--(5);
	\draw (4)--(6);
	\begin{scope}[red]
	\draw (1)--(4)--(2);
	\draw (5)--(3)--(6);
	\end{scope};
	
	\draw (11,5) node (1)  {};
	\draw (13,3) node (2)  {};
	\draw (13,7) node (3)  {};
	\draw (17,3) node (4)  {};
	\draw (17,7) node (5)  {};
	\draw (19,5) node (6)  {};

	\draw (3)--(1)--(2)--(3)--(5)--(4)--(6)--(5);
	\draw (3)--(4);
	\draw (4)--(2)--(5);
	\begin{scope}[red]
	\draw (4)--(1)--(5);
	\draw (2)--(6)--(3);
	\end{scope};

	\draw (22,5) node (1b)  {};
	\draw (24,3) node (2b)  {};
	\draw (24,7) node (3b)  {};
	\draw (28,3) node (4b)  {};
	\draw (28,7) node (5b)  {};
	\draw (30,5) node (6b)  {};

	\draw (3b)--(1b)--(2b)--(3b)--(5b)--(4b)--(6b)--(5b);
	\draw (3b)--(4b);
	\draw (2b)--(5b);
	\begin{scope}[red]
	\draw (4b)--(1b)--(5b);
	\draw (2b)--(6b)--(3b);
	\end{scope}; 
	
	\coordinate[label=left:{$H_1$}] (h1) at (5,1.5);
	
	\coordinate[label=left:{$H_2$}] (h2) at (16,1.5);
	
	\coordinate[label=left:{$H_3$}] (h3) at (27,1.5);

	\end{tikzpicture}
	
	\caption{Edges added to forbidden graphs, in red to convert them to chordal (6, 3).}
	\label{fig:overview}
\end{figure}
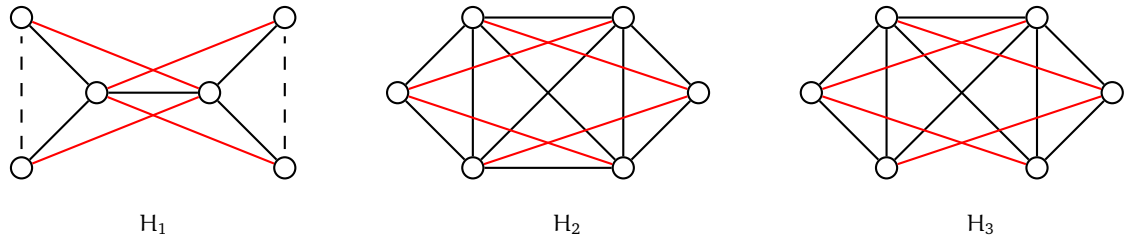

 A system for marking the edges added to Babel's forbidden subgraphs to convert them to the extended chordal (6, 3) graphs is needed. To each node that is part of an added edge a tree graph is attached.  This tree can encode information specific to the node ie. which type of forbidden subgraph it is part of what position it holds in the subgraph (see fig. 4, 5).  After a graph's forbidden subgraphs have been found and the edges added to them and marked new forbidden subgraphs may be created.  So a second round of finding forbidden subgraphs adding edges and marking them will be necessary.  To mark the second set of added edges the height of the marking trees has to be increased, to create a second level.  There may be forbidden subgraphs that contain one or two root nodes of marking trees.  Edges will not be added to these nodes (see fig. 6).  After the second round of finding and marking a third round may be necessary.  Since the graph can have at most $n(n-1)/2$ edges this processes has to stop.
 
  \begin{figure}[H]
 	\centering\small
 	
 	\begin{tikzpicture}[thick, scale=0.5]
 	
 	\tikzstyle{every node}=[circle,draw=black,fill=white,thick,minimum size=8pt,
 	inner sep=0pt]
 	
 	\draw (20,3) node (1b) {};
 	\draw (20,5) node (2b) {};
 	\draw (21,4) node (3b) {};
 	\draw (23,4) node (4b)[label=right:node type 2] {};
 	\draw (24,3) node (5b)[label=right:node type 1] {};
 	\draw (24,5) node (6b) {};
 	
 	\draw (1b)--(3b)--(4b)--(6b) (2b)--(3b) (4b)--(5b);
 	
 	\coordinate [label=left:$H_{3}$] (H3) at (23,14);
 	
 	\draw (20,10) node (1c)  {};
 	\draw (21,9) node (2c)  {};
 	\draw (21,11) node (3c)  {};
 	\draw (23,9) node (4c)[label=right:node type 2] {};
 	\draw (23,11) node (5c)  {};
 	\draw (24,10) node (6c)[label=right:node type 1] {};
 	
 	\draw (1c)--(3c)--(5c)--(6c) (1c)--(2c)--(4c)--(6c) (3c)--(4c) (2c)--(5c);
 	
 	\coordinate [label=left:$H_{2}$] (H2) at (23,8);
 	
 	\draw (20,16) node (1d)  {};
 	\draw (21,15) node (2d)  {};
 	\draw (21,17) node (3d)  {};
 	\draw (23,15) node (4d)[label=right:node type 2] {};
 	\draw (23,17) node (5d)  {};
 	\draw (24,16) node (6d)[label=right:node type 1] {};
 	
 	\draw (1d)--(3d)--(5d)--(6d) (1d)--(2d) (4d)--(6d) (3d)--(4d) (2d)--(5d);
 	
 	\coordinate [label=left:$H_{1}$] (H1) at (23,2);

 	\draw (2,34) node (33)  {};
 	\draw (2,33) node (32)  {};
 	\draw (2,32) node (31)  {};
 	\draw (2,31) node (30) [label={[xshift=2.0cm,  yshift= -1.8cm]Round Three}] {};
 	\draw (2,30) node (29)  {};
 	\draw (2,29) node (28)  {};
 	\draw (2,28) node (27)  {};
 	\draw (2,27) node (26)  {};
 	\draw (2,26) node (25)  {};
 	\draw (2,25) node (24)  {};
 	\draw (2,24) node (23)  {};
 	\draw (2,23) node (22)  {};
 	\draw (2,22) node (21)  {};
 	\draw (2,21) node (20)  {};
 	\draw (2,20) node (19)  {};
 	\draw (2,19) node (18)  {};
 	\draw (2,18) node (17) {};
 	\draw (2,17) node (16) {};
 	\draw (2,16) node (15) [label={[xshift=2.0cm,  yshift= -1.2cm]Round Two}] {};
 	\draw (2,15) node (14)  {};
 	\draw (2,14) node (13)  {};
 	\draw (2,13) node (12)  {};
 	\draw (2,12) node (11)  {};
 	\draw (2,11) node (10)  {};
 	\draw (2,10) node (9)  {};
 	\draw (2,9) node (8)  {};
 	\draw (2,8) node (7)  {};
 	\draw (2,7) node (6)  {};
 	\draw (2,6) node (5)  {};
 	\draw (2,5) node (4)  {};
 	\draw (2,4) node (3)  {};
 	\draw (2,3) node (2)  {};
 	\draw (2,2) node (1)  {};

 	\draw (3.5,33) node (16a)  {};
 	\draw (6,33) node (16b)  {};
 	\draw (3.5,31) node (15a)  {};
 	\draw (6,31) node (15b)  {};
 	\draw (3.5,29) node (14a)  {};
 	\draw (6,29) node (14b) [label=right:number of edges $H_{3}$ node type 2] {};
 	\draw (3.5,27) node (13a)  {};
 	\draw (6,27) node (13b) [label=right:number of edges $H_{3}$ node type 1] {};
 	\draw (3.5,25) node (12a)  {};
 	\draw (6,25) node (12b) [label=right:number of edges $H_{2}$ node type 2] {};
 	\draw (3.5,23) node (11a)  {};
 	\draw (6,23) node (11b) [label=right:number of edges $H_{2}$ node type 1] {};
 	\draw (3.5,21) node (10a)  {};
 	\draw (6,21) node (10b) [label=right:number of edges $H_{1}$ node type 2] {};
 	\draw (3.5,19) node (9a)  {};
 	\draw (6,19) node (9b) [label=right:edges $H_{1}$ between node type 1] {};
 	\draw (3.5,17) node (8a) {};
 	\draw (6,17) node (8b) [label=right:number of edges $H_{1}$ node type 1] {};
 	\draw (3.5,15) node (7a) {};
 	\draw (6,15) node (7b) [label=right:number of edges $H_{3}$ node type 2] {};
 	\draw (3.5,13) node (6a)  {};
 	\draw (6,13) node (6b) [label=right:number of edges $H_{3}$ node type 1] {};
 	\draw (3.5,11) node (5a)  {};
 	\draw (6,11) node (5b) [label=right:number of edges $H_{2}$ node type 2] {};
 	\draw (3.5,9) node (4a)  {};
 	\draw (6,9) node (4b) [label=right:number of edges $H_{2}$ node type 1] {};
 	\draw (3.5,7) node (3a)  {};
 	\draw (6,7) node (3b) [label=right:number of edges $H_{1}$ node type 2] {};
 	\draw (3.5,5) node (2a)  {};
 	\draw (6,5) node (2b) [label=right:edges $H_{1}$ between node type 1] {};
 	\draw (3.5,3) node (1a)  {};
 	\draw (6,3) node (1b) [label=right:number of edges $H_{1}$ node type 1] {};
 	
 	\begin{scope}[green]
 	
 	\draw (1)--(2)--(3)--(4)--(5)--(6)--(7)--(8)--(9)--(10)--(11)--(12)--(13)--(14)--(15)--(16)--(17)--(18)--(19)--(20)--(21)--(22)--(23)--(24)--(25)--(26)--(27)--(28)--(29)--(30)--(31)--(32)--(33);
 	
 	\draw [thick] (2.5,3)--(3,3);
 	\draw [thick,dash pattern={on 7pt off 7pt}] (4,3) -- (5.5,3); 
 	\draw [thick] (2.5,5)--(3,5);
 	\draw [thick,dash pattern={on 7pt off 7pt}] (4,5) -- (5.5,5);
 	\draw [thick] (2.5,7)--(3,7);
 	\draw [thick,dash pattern={on 7pt off 7pt}] (4,7) -- (5.5,7);
 	\draw [thick] (2.5,9)--(3,9);
 	\draw [thick,dash pattern={on 7pt off 7pt}] (4,9) -- (5.5,9);
 	\draw [thick] (2.5,11)--(3,11);
 	\draw [thick,dash pattern={on 7pt off 7pt}] (4,11) -- (5.5,11);
 	\draw [thick] (2.5,13)--(3,13);
 	\draw [thick,dash pattern={on 7pt off 7pt}] (4,13) -- (5.5,13);
 	\draw [thick] (2.5,15)--(3,15);
 	\draw [thick,dash pattern={on 7pt off 7pt}] (4,15) -- (5.5,15);
 	\draw [thick] (2.5,17)--(3,17);
 	\draw [thick,dash pattern={on 7pt off 7pt}] (4,17) -- (5.5,17);
 	\draw [thick] (2.5,19)--(3,19);
 	\draw [thick,dash pattern={on 7pt off 7pt}] (4,19) -- (5.5,19);
 	\draw [thick] (2.5,21)--(3,21);
 	\draw [thick,dash pattern={on 7pt off 7pt}] (4,21) -- (5.5,21);
 	\draw [thick] (2.5,23)--(3,23);
 	\draw [thick,dash pattern={on 7pt off 7pt}] (4,23) -- (5.5,23);
 	\draw [thick] (2.5,25)--(3,25);
 	\draw [thick,dash pattern={on 7pt off 7pt}] (4,25) -- (5.5,25);
 	\draw [thick] (2.5,27)--(3,27);
 	\draw [thick,dash pattern={on 7pt off 7pt}] (4,27) -- (5.5,27);
 	\draw [thick] (2.5,29)--(3,29);
 	\draw [thick,dash pattern={on 7pt off 7pt}] (4,29) -- (5.5,29);
 	\draw [thick] (2.5,31)--(3,31);
 	\draw [thick,dash pattern={on 7pt off 7pt}] (4,31) -- (5.5,31);
 	\draw [thick] (2.5,33)--(3,33);
 	\draw [thick,dash pattern={on 7pt off 7pt}] (4,33) -- (5.5,33);
 	
 	\end{scope}
 	
 	\draw [thick,dash pattern={on 3pt off 3pt}] (20,3.5) -- (20,4.5);
 	\draw [thick,dash pattern={on 3pt off 3pt}] (24,3.5) -- (24,4.5);

 	\end{tikzpicture}
 	
 	\caption{Tree graph for marking end nodes of edges added to eliminate forbidden graphs.}
 	\label{fig:overview}
 \end{figure}
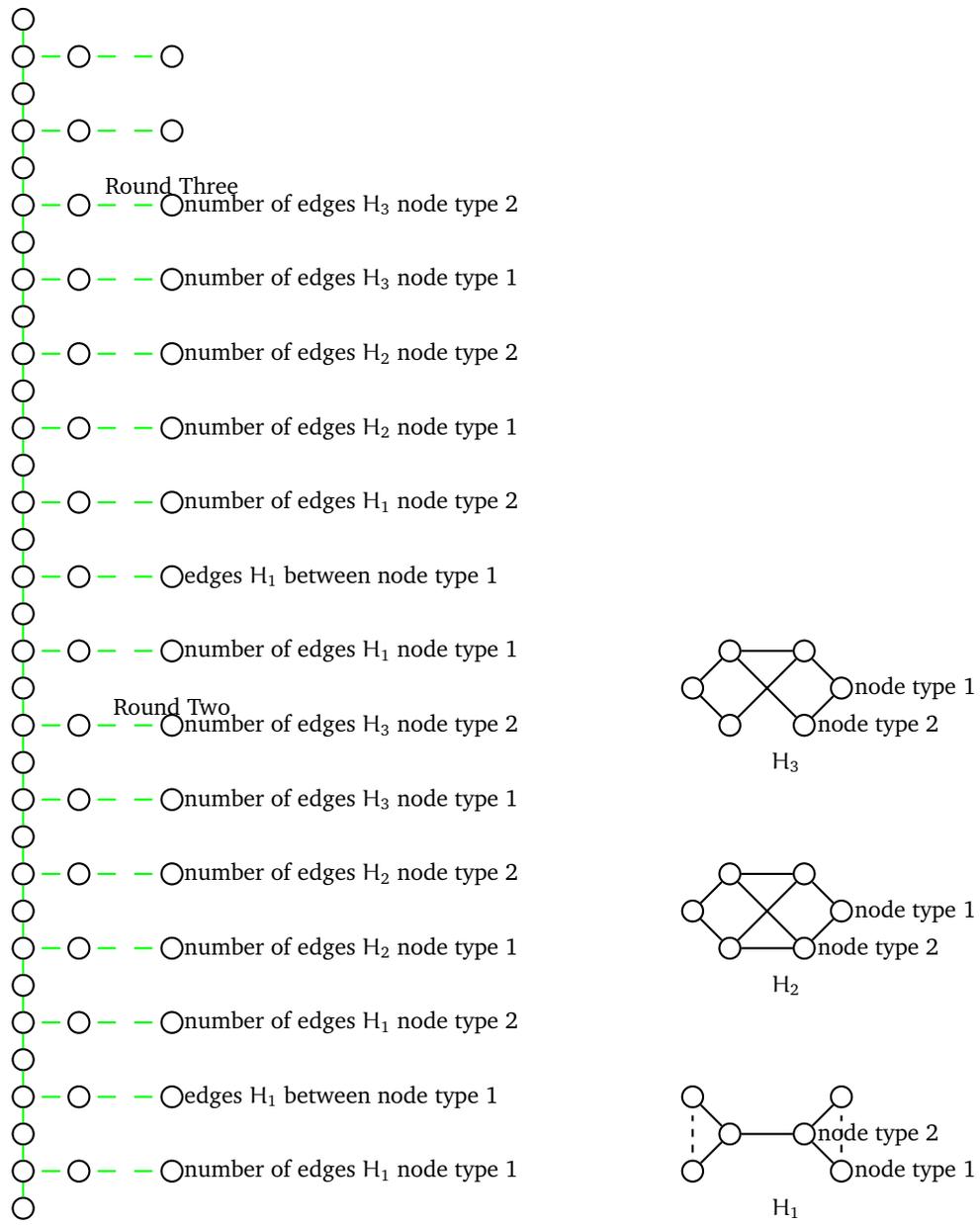

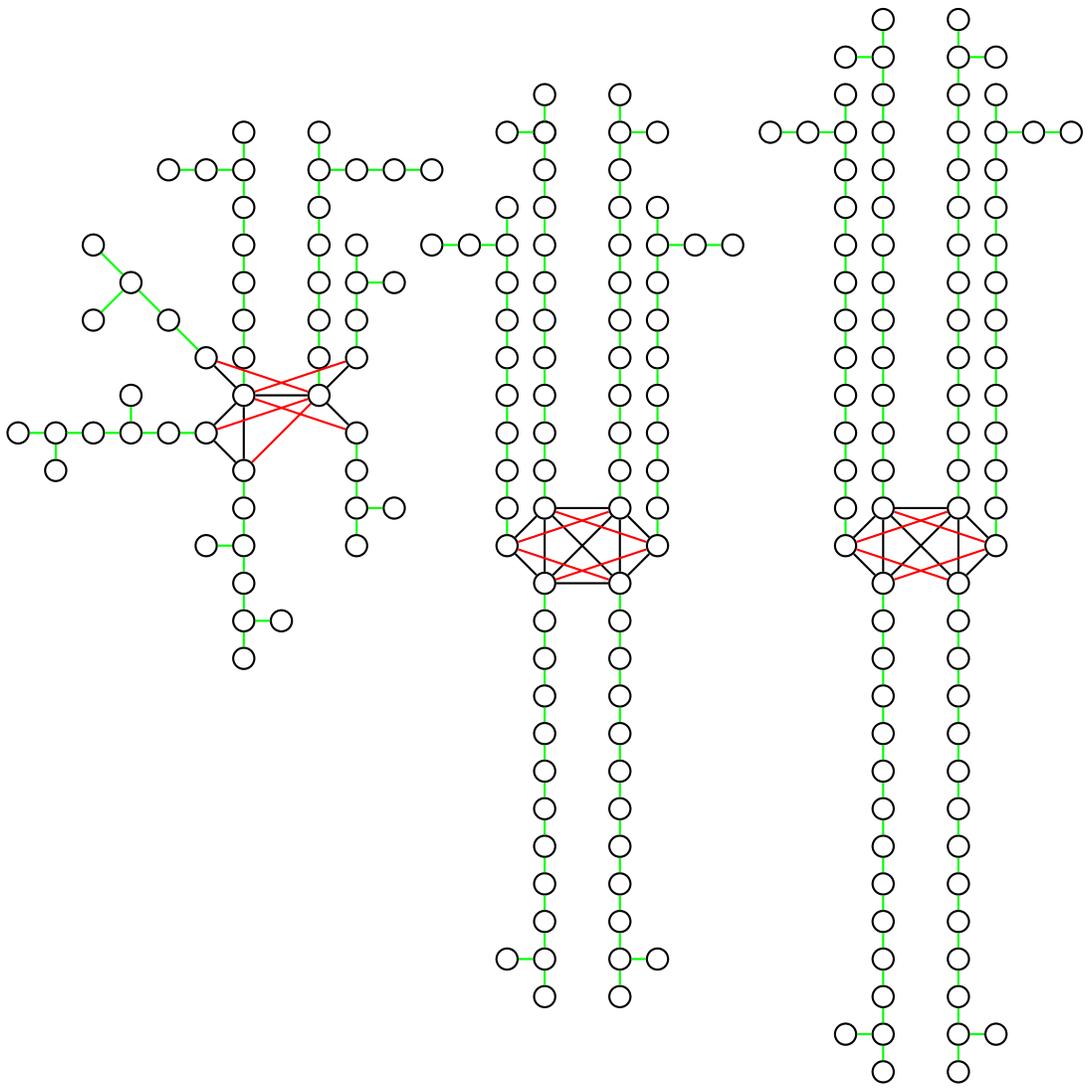
\begin{figure}[H]
	\centering\small
	
	\begin{tikzpicture}[thick, scale=0.5]
	
	\tikzstyle{every node}=[circle,draw=black,fill=white,thick,minimum size=8pt,
	inner sep=0pt]
	
	\draw (7,19) node (1a)  {};
	\draw (7,21) node (2a)  {};
	\draw (8,20) node (3a)  {};
	\draw (10,20) node (4a)  {};
	\draw (11,19) node (5a)  {};
	\draw (11,21) node (6a)  {};
	\draw (8,18) node (7x)  {};
	
	\draw (6,22) node (7a)  {};
	\draw (5,23) node (8a)  {};
	\draw (4,22) node (9a)  {};
	\draw (4,24) node (10a)  {};
	
	\draw (8,21) node (11a)  {};
	\draw (8,22) node (12a)  {};
	\draw (8,23) node (13a)  {};
	\draw (8,24) node (14a)  {};
	\draw (8,25) node (15a)  {};
	\draw (8,26) node (16a)  {};
	\draw (7,26) node (17a)  {};
	\draw (6,26) node (17x)  {};
	\draw (8,27) node (18x)  {};
	
	\draw (10,21) node (18a)  {};
	\draw (10,22) node (19a)  {};
	\draw (10,23) node (20a)  {};
	\draw (10,24) node (21a)  {};
	\draw (10,25) node (22a)  {};
	\draw (10,26) node (23a)  {};
	\draw (11,26) node (24a)  {};
	\draw (12,26) node (24x)  {};
	\draw (13,26) node (24xx)  {};
	\draw (10,27) node (25x)  {};

	\draw (11,22) node (25a)  {};
	\draw (11,23) node (26a)  {};
	\draw (12,23) node (27a)  {};
	\draw (11,24) node (28a)  {};
	
	\draw (2,19) node (29x)  {};
	\draw (3,18) node (30x)  {};
	\draw (3,19) node (31x)  {};
	\draw (4,19) node (29a)  {};
	\draw (5,19) node (30a)  {};
	\draw (5,20) node (31a)  {};
	\draw (6,19) node (32a)  {};
	
	\draw (8,17) node (11x)  {};
	\draw (8,16) node (12x)  {};
	\draw (7,16) node (13x)  {};
	\draw (8,15) node (14x)  {};
	\draw (8,14) node (15x)  {};
	\draw (9,14) node (16x)  {};
	\draw (8,13) node (17x2)  {};
	
	\draw (11,16) node (33a)  {};
	\draw (11,17) node (34a)  {};
	\draw (12,17) node (35a)  {};
	\draw (11,18) node (36a)  {};
	
	\draw (1a)--(3a)--(2a);
	\draw (1a)--(7x)--(3a)--(4a)--(5a);
	\draw (4a)--(6a);
	
	\begin{scope}[green]
	\draw (2a)--(7a)--(8a)--(10a);
	\draw (8a)--(9a);
	\draw (3a)--(11a)--(12a)--(13a)--(14a)--(15a)--(16a)--(18x);
	\draw (16a)--(17a)--(17x);
	\draw (4a)--(18a)--(19a)--(20a)--(21a)--(22a)--(23a)--(25x);
	\draw(23a)--(24a)--(24x)--(24xx);                                                
	\draw (6a)--(25a)--(26a)--(28a);
	\draw (26a)--(27a);
	\draw (1a)--(32a)--(30a)--(29a)--(31x)--(29x);
	\draw (30x)--(31x);
	\draw (30a)--(31a);
	\draw (5a)--(36a)--(34a)--(33a);
	\draw (34a)--(35a);
	\draw (7x)--(11x)--(12x)--(14x)--(15x)--(17x2);
	\draw (13x)--(12x);
	\draw (16x)--(15x);
	\end{scope}
	
	\begin{scope}[red]
	\draw (2a)--(4a)--(1a);
	\draw (5a)--(3a)--(6a);
	\draw (7x)--(4a);
	\end{scope}
	
	\draw (15,16) node (1b)  {};
	\draw (16,15) node (2b) {};
	\draw (16,17) node (3b) {};
	\draw (18,15) node (4b) {};
	\draw (18,17) node (5b) {};
	\draw (19,16) node (6b) {};
	
	\draw (15,17) node (7y)  {};
	\draw (15,18) node (7b)  {};
	\draw (15,19) node (8b)  {};
	\draw (15,20) node (9b)  {};
	\draw (15,21) node (10b)  {};
	\draw (15,22) node (11b)  {};
	\draw (15,23) node (12b)  {};
	\draw (13,24) node (13bb)  {};
	\draw (14,24) node (13b)  {};
	\draw (15,24) node (14b)  {};
	\draw (15,25) node (14y)  {};
	
	\draw (16,18) node (15b)  {};
	\draw (16,19) node (16b)  {};
	\draw (16,20) node (17b)  {};
	\draw (16,21) node (18b)  {};
	\draw (16,22) node (19b)  {};
	\draw (16,23) node (20b)  {};
	\draw (16,24) node (21b)  {};
	\draw (16,25) node (22b)  {};
	\draw (16,26) node (23b)  {};
	\draw (15,27) node (24b)  {};
	\draw (16,27) node (25b)  {};
	\draw (16,27) node (26b)  {};
	\draw (16,28) node (26y)  {};
	
	\draw (18,18) node (27b)  {};
	\draw (18,19) node (28b)  {};
	\draw (18,20) node (29b)  {};
	\draw (18,21) node (30b)  {};
	\draw (18,22) node (31b)  {};
	\draw (18,23) node (32b)  {};
	\draw (18,24) node (33b)  {};
	\draw (18,25) node (34b)  {};
	\draw (18,26) node (35b)  {};
	\draw (19,27) node (36b)  {};
	\draw (18,27) node (38b)  {};
	\draw (18,28) node (38y)  {};
	
	\draw (19,17) node (39y)  {};
	\draw (19,18) node (39b)  {};
	\draw (19,19) node (40b)  {};
	\draw (19,20) node (41b)  {};
	\draw (19,21) node (42b)  {};
	\draw (19,22) node (43b)  {};
	\draw (19,23) node (44b)  {};
	\draw (21,24) node (45bb)  {};
	\draw (20,24) node (45b)  {};
	\draw (19,24) node (46b)  {};
	\draw (19,25) node (46y)  {};
	
	\draw (16,14) node (45d)  {};
	\draw (16,13) node (46d)  {};
	\draw (16,12) node (47y)  {};
	\draw (16,11) node (47b)  {};
	\draw (16,10) node (48b)  {};
	\draw (16,9) node (49b)  {};
	\draw (16,8) node (50b)  {};
	\draw (16,7) node (51b)  {};
	\draw (15,5) node (52b)  {};
	\draw (16,6) node (53b)  {};
	\draw (16,5) node (54b)  {};
	\draw (16,4) node (54y)  {};
	
	\draw (18,14) node (53d)  {};
	\draw (18,13) node (54d)  {};
	\draw (18,12) node (55y)  {};
	\draw (18,11) node (55b)  {};
	\draw (18,10) node (56b)  {};
	\draw (18,9) node (57b)  {};
	\draw (18,8) node (58b)  {};
	\draw (18,7) node (59b)  {};
	\draw (19,5) node (60b)  {};
	\draw (18,6) node (61b)  {};
	\draw (18,5) node (62b)  {};
	\draw (18,4) node (62y)  {};
	
	\draw (4b)--(2b)--(1b)--(3b)--(2b)--(5b);
	\draw (5b)--(3b)--(4b)--(5b);
	\draw (4b)--(6b)--(5b);
	
	\begin{scope}[green]
	\draw (1b)--(7y)--(7b)--(8b)--(9b)--(10b)--(11b)--(12b)--(14b)--(14y);
	\draw (13bb)--(13b)--(14b);
	\draw (3b)--(15b)--(16b)--(17b)--(18b)--(19b)--(20b)--(21b)--(22b)--(23b)--(26b)--(26y);
	\draw (26b)--(24b)--(25b);
	\draw (5b)--(27b)--(28b)--(29b)--(30b)--(31b)--(32b)--(33b)--(34b)--(35b)--(38b)--(38y);
	\draw (38b)--(36b);
	\draw (6b)--(39y)--(39b)--(40b)--(41b)--(42b)--(43b)--(44b)--(46b)--(46y);
	\draw (45bb)--(45b)--(46b);
	\draw (2b)--(45d)--(46d)--(47y)--(47b)--(48b)--(49b)--(50b)--(51b)--(53b)--(54b)--(54y);
	\draw (52b)--(54b);
	\draw (4b)--(53d)--(54d)--(55y)--(55b)--(56b)--(57b)--(58b)--(59b)--(61b)--(62b)--(62y);
	\draw (60b)--(62b);
	\end{scope}
	
	\begin{scope}[red]
	\draw (5b)--(1b)--(4b);
	\draw (2b)--(6b)--(3b);
	\end{scope}

	\draw (25,15) node (3c)  {};
	\draw (25,17) node (2c)  {};
	\draw (24,16) node (1c)  {};
	\draw (28,16) node (6c)  {};
	\draw (27,15) node (5c)  {};
	\draw (27,17) node (4c)  {};
	
	\draw (24,17) node (7z)  {};
	\draw (24,18) node (7c)  {};
	\draw (24,19) node (8c)  {};
	\draw (24,20) node (9c)  {};
	\draw (24,21) node (10c)  {};
	\draw (24,22) node (11c)  {};
	\draw (24,23) node (12c)  {};
	\draw (24,24) node (13c)  {};
	\draw (24,25) node (14c)  {};
	\draw (24,26) node (15c)  {};
	\draw (24,27) node (16c)  {};
	\draw (23,27) node (17c)  {};
	\draw (22,27) node (17cc)  {};
	\draw (24,28) node (18c)  {};
	
	\draw (25,18) node (19c)  {};
	\draw (25,19) node (20c)  {};
	\draw (25,20) node (21c)  {};
	\draw (25,21) node (22c)  {};
	\draw (25,22) node (23c)  {};
	\draw (25,23) node (24c)  {};
	\draw (25,24) node (25c)  {};
	\draw (25,25) node (26c)  {};
	\draw (25,26) node (27c)  {};
	\draw (25,27) node (28c)  {};
	\draw (25,28) node (29c)  {};
	\draw (25,29) node (30c)  {};
	\draw (24,29) node (31c)  {};
	\draw (25,30) node (33c)  {};
	
	\draw (27,18) node (34c)  {};
	\draw (27,19) node (35c)  {};
	\draw (27,20) node (36c)  {};
	\draw (27,21) node (37c)  {};
	\draw (27,22) node (38c)  {};
	\draw (27,23) node (39c)  {};
	\draw (27,24) node (40c)  {};
	\draw (27,25) node (41c)  {};
	\draw (27,26) node (42c)  {};
	\draw (27,27) node (43c)  {};
	\draw (27,28) node (44c)  {};
	\draw (27,29) node (45c)  {};
	\draw (28,29) node (46c)  {};
	\draw (39,29) node (47c)  {};
	\draw (27,30) node (48c)  {};
	
	\draw (28,17) node (49z)  {};
	\draw (28,18) node (49c)  {};
	\draw (28,19) node (50c)  {};
	\draw (28,20) node (51c)  {};
	\draw (28,21) node (52c)  {};
	\draw (28,22) node (53c)  {};
	\draw (28,23) node (54c)  {};
	\draw (28,24) node (55c)  {};
	\draw (28,25) node (56c)  {};
	\draw (28,26) node (57c)  {};
	\draw (28,27) node (58c)  {};
	\draw (29,27) node (59c)  {};
	\draw (30,27) node (59cc)  {};
	\draw (28,28) node (60c)  {};
	
	\draw (25,14) node (61c)  {};
	\draw (25,13) node (62c)  {};
	\draw (25,12) node (63c)  {};
	\draw (25,11) node (64c)  {};
	\draw (25,10) node (65c)  {};
	\draw (25,9) node (66c)  {};
	\draw (25,8) node (67c)  {};
	\draw (25,7) node (68c)  {};
	\draw (25,6) node (69c)  {};
	\draw (25,5) node (70c)  {};
	\draw (24,3) node (72cc)  {};
	\draw (25,4) node (71c)  {};
	\draw (25,3) node (72c)  {};
	\draw (25,2) node (73d)  {};
	
	\draw (27,14) node (73c)  {};
	\draw (27,13) node (74c)  {};
	\draw (27,12) node (75c)  {};
	\draw (27,11) node (76c)  {};
	\draw (27,10) node (77c)  {};
	\draw (27,9) node (78c)  {};
	\draw (27,8) node (79c)  {};
	\draw (27,7) node (80c)  {};
	\draw (27,6) node (81c)  {};
	\draw (28,3) node (82c)  {};
	\draw (27,5) node (83c)  {};
	\draw (27,4) node (84c)  {};
	\draw (27,3) node (85c)  {};
	\draw (27,2) node (86c)  {};
	
	\draw (2c)--(3c)--(1c)--(2c);
	\draw (4c)--(6c)--(5c)--(4c);
	\draw (4c)--(2c)--(5c);
	\draw (4c)--(3c);
	
	\begin{scope}[red]
	\draw (5c)--(1c)--(4c);
	\draw (2c)--(6c)--(3c);
	\end{scope}
	
	\begin{scope}[green]
	\draw (1c)--(7z)--(7c)--(8c)--(9c)--(10c)--(11c)--(12c)--(13c)--(14c)--(15c)--(16c)--(18c);
	\draw (16c)--(17c)--(17cc);
	\draw (2c)--(19c)--(20c)--(21c)--(22c)--(23c)--(24c)--(25c)--(26c)--(27c)--(28c)--(29c)--(30c)--(33c);
	\draw (30c)--(31c);
	\draw (4c)--(34c)--(35c)--(36c)--(37c)--(38c)-- (39c)--(40c)--(41c)--(42c)--(43c)--(44c)--(45c)--(46c)(45c)--(48c); 
	\draw (6c)--(49z)--(49c)--(50c)--(51c)--(52c)--(53c)--(54c)--(55c)--(56c)--(57c)--(58c)--(59c)--(59cc)(58c)--(60c);
	\draw (3c)--(61c)--(62c)--(63c)--(64c)--(65c)--(66c)--(67c)--(68c)--(69c)--(70c)--(71c)--(72c)--(73d)(72c)--(72cc);
	\draw (5c)--(73c)--(74c)--(75c)--(76c)--(77c)--(78c)--(79c)--(80c)--(81c)--(83c)--(84c)--(85c)--(86c)(85c)--(82c);
	\end{scope}
	
	\end{tikzpicture}
	
	\caption{Examples of round 1 marked forbidden graphs}
	\label{fig:overview}
\end{figure}

  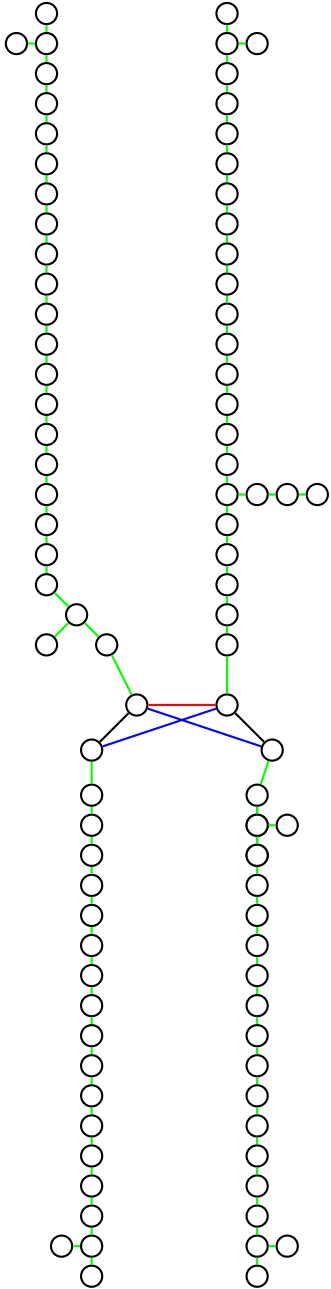
\begin{figure}[H]
	\centering\small
	
	\begin{tikzpicture}[thick, scale=0.4]
	
	\tikzstyle{every node}=[circle,draw=black,fill=white,thick,minimum size=8pt,
	inner sep=0pt]
	
	\draw (7,20) node (2a)  {};
	\draw (5.5,18.5) node (3a)  {};
	\draw (10,20) node (4a)  {};
	\draw (11.5,18.5) node (5a)  {};
	
	\draw (6,22) node (7a)  {};
	\draw (5,23) node (8a)  {};
	\draw (4,22) node (9a)  {};
	\draw (4,24) node (10a)  {};
	\draw (4,25) node (11a)  {};
	\draw (4,26) node (12a)  {};
	\draw (4,27) node (13a)  {};
	\draw (4,28) node (14a)  {};
	\draw (4,29) node (15a)  {};
	\draw (4,30) node (16a)  {};
	\draw (4,31) node (17a)  {};
	\draw (4,32) node (18a)  {};
	\draw (4,33) node (19a)  {};
	\draw (4,34) node (20a)  {};
	\draw (4,35) node (21a)  {};
	\draw (4,36) node (22a)  {};
	\draw (4,37) node (23a)  {};
	\draw (4,38) node (24a)  {};
	\draw (4,39) node (25a)  {};
	\draw (4,40) node (26a)  {};
	\draw (4,41) node (27a)  {};
	\draw (4,42) node (28a)  {};
	\draw (4,43) node (29a)  {};
	\draw (3,42) node (30a)  {};

	\draw (5.5,17) node (11x)  {};
	\draw (5.5,16) node (12x)  {};
	\draw (5.5,15) node (13x)  {};
	\draw (5.5,14) node (14x)  {};
	\draw (5.5,13) node (15x)  {};
	\draw (5.5,12) node (16x)  {};
	\draw (5.5,11) node (17x)  {};
	\draw (5.5,10) node (18x)  {};
	\draw (5.5,9) node (19x)  {};
	\draw (5.5,8) node (20x)  {};
	\draw (5.5,7) node (21x)  {};
	\draw (5.5,6) node (22x)  {};
	\draw (5.5,5) node (23x)  {};
	\draw (5.5,4) node (24x)  {};
	\draw (5.5,3) node (25x)  {};
	\draw (5.5,2) node (26x)  {};
	\draw (5.5,1) node (27x)  {};
	\draw (4.5,2) node (28x)  {};

	\draw (10,22) node (16y)  {};
	\draw (10,23) node (17y)  {};
	\draw (10,24) node (18y)  {};
	\draw (10,25) node (19y)  {};
	\draw (10,26) node (20y)  {};
	\draw (10,27) node (21y)  {};
	\draw (11,27) node (22y)  {};
	\draw (12,27) node (23y)  {};
	\draw (13,27) node (24y)  {};
	\draw (10,28) node (25y)  {};
	\draw (10,29) node (26y)  {};
	\draw (10,30) node (27y)  {};
	\draw (10,31) node (28y)  {};
	\draw (10,32) node (29y)  {};
	\draw (10,33) node (30y)  {};
	\draw (10,34) node (31y)  {};
	\draw (10,35) node (32y)  {};
	\draw (10,36) node (33y)  {};
	\draw (10,37) node (34y)  {};
	\draw (10,38) node (35y)  {};
	\draw (10,39) node (36y)  {};
	\draw (10,40) node (37y)  {};
	\draw (10,41) node (38y)  {};
	\draw (10,42) node (39y)  {};
	\draw (10,43) node (40y)  {};
	\draw (11,42) node (41y)  {};

	\draw (11,15) node (33a)  {};
	\draw (11,16) node (34a)  {};
	\draw (12,16) node (35a)  {};
	\draw (11,17) node (36a)  {};
	\draw (11,16) node (37a)  {};
	\draw (11,15) node (38a)  {};
	\draw (11,14) node (39a)  {};
	\draw (11,13) node (40a)  {};
	\draw (11,12) node (41a)  {};
	\draw (11,11) node (42a)  {};
	\draw (11,10) node (43a)  {};
	\draw (11,9) node (44a)  {};
	\draw (11,8) node (45a)  {};
	\draw (11,7) node (46a)  {};
	\draw (11,6) node (47a)  {};
	\draw (11,5) node (48a)  {};
	\draw (11,4) node (49a)  {};
	\draw (11,3) node (50a)  {};
	\draw (11,2) node (51a)  {};
	\draw (11,1) node (52a)  {};
	\draw (12,2) node (53a)  {};

	\draw (3a)--(2a);
	\draw (4a)--(5a);
	\begin{scope}[blue]
	\draw (2a)--(5a);
	\draw (3a)--(4a);
	\end{scope}
	
	\begin{scope}[green]
	\draw (2a)--(7a)--(8a)--(10a)--(11a)--(12a)--(13a)--(14a)--(15a)--(16a)--(17a)--(18a)--(19a)--(20a)--(21a)--(22a)--(23a)--(24a)--(25a)--(26a)--(27a)--(28a)--(29a);
	\draw (8a)--(9a)(28a)--(30a);
	\draw (4a)--(16y)--(17y)--(18y)--(19y)--(20y)--(21y)--(25y)--(26y)--(27y)--(28y)--(29y)--(30y)--(31y)--(32y)--(33y)--(34y)--(35y)--(36y)--(37y)--(38y)--(39y)--(40y) (41y)--(39y);
	\draw (21y)--(22y)--(23y)--(24y);
	\draw (3a)--(11x)--(12x)--(13x)--(14x)--(15x)--(16x)--(17x)--(18x)--(19x)--(20x)--(21x)--(22x)--(23x)--(24x)--(25x)--(26x)--(27x) (26x)--(28x);
	\draw (36a)--(37a)--(38a)--(39a)--(40a)--(41a)--(42a)--(43a)--(44a)--(45a)--(46a)--(47a)--(48a)--(49a)--(50a)--(51a)--(52a) (51a)--(53a);
	
	\draw (5a)--(36a)--(34a)--(33a);
	\draw (34a)--(35a);
	\end{scope}
	
	\begin{scope}[red]
	\draw (2a)--(4a);
	\end{scope}

	\end{tikzpicture}
	
	\caption{Example of round 2 marked forbidden graph. Second round added edges in blue.  First round added edge in red.  Notice that edges are not added to the tree nodes. }
	\label{fig:overview}
\end{figure}

\section{Isomorphism Testing}\label{sec:final}
To test whether two graphs are isomorphic first test if graphs are chordal. If the graphs are not chordal apply Booth's reduction to them.  If the graphs are chordal test whether they contain any of the forbidden subgraphs.  If the graphs contain forbidden subgraphs eliminate them by adding edges to them and marking them repeat this elimination process till the graphs are free of forbidden subgraphs.  There are now two extended chordal (6, 3) graphs which may be tested as in Babel.\par
 An alternate procedure that does not use marking trees can be done as follows.  Create the two extended chordal (6, 3) graphs as above using booth reduction and adding edges to eliminate forbidden graphs as needed.  Test the graphs as in Babel if the automorphism partition is trivial the graphs are not isomorphic.  If the partition is not trivial find the corresponding vertices in the two graphs from the simplicial partition align the graphs and see if the original edges in the two graphs match.  If the edges do not match the graphs are not isomorphic.


\begin{thebibliography}{10}
	
	\bibitem{babel-95-chordal-(6, 3)-isomorphism}
	L. Babel.
	\newblock Isomorphism of chordal (6, 3) graphs
	\newblock Computing 54(4): 303-316 (1995)

	\bibitem{babel-95-graph-p4s-isomorphism}
	L. Babel, S. Olariu.
	\newblock On the isomorphism of graphs with few P4S.
	\newblock  Workshop on Graph-Theoretic Concepts in Comp. Sci. WG '95, M. Nagl, ed., Lecture Notes in Comput. Sci, 1017 (1995), 24-36
	
	\bibitem{babel-96-isomorphism-directed-path-graphs}
	L. Babel, I.N.Ponomarenko, and G. Tinhofer.
	\newblock The Isomorphism Problem For Directed Path Graphs and For Rooted Directed Path Graphs
	\newblock {\em J. Algorithms} 21 (1996) 542–-564.
	
	\bibitem{booth-79-polynomially-equivalent-graph-isomorphism}
    Booth, Kellogg S.; Colbourn, C. J.
	\newblock Problems polynomially equivalent to graph isomorphism.
	\newblock Technical Report, No. CS-77-04, Computer Science Department, University of Waterloo 1979.
	
	
\end{thebibliography}
\end{document}